\documentclass[twocolumn,amsmath,amssymb]{revtex4}

\usepackage{epsfig}
\usepackage{array}
\usepackage{multirow}

\begin{document}

\title{Control of drop positioning using chemical patterning}

\author{A.~Dupuis$^1$}
\author{J.~L\'{e}opold\`{e}s$^2$}
\author{D.G.~Bucknall$^2$}
\author{J.M.~Yeomans$^1$}
\affiliation{$^1$ The Rudolf Peierls Centre for Theoretical Physics, University of Oxford, 1 Keble Road, Oxford OX1 3NP, UK\\
$^2$ Department of Materials, Oxford University, Oxford OX1 3PH, UK.}

\date{\today}

\begin{abstract}
We explore how chemical patterning on surfaces can be used to control
drop wetting. Both numerical and experimental results are presented to
show how the dynamic pathway and equilibrium shape of the drops are
altered by a hydrophobic grid. The grid proves a successful way of
confining drops and we show that it can be used to alleviate {\it
mottle}, a degradation in image quality which results from uneven drop
coalescence due to randomness in the positions of the drops within the
jetted array.
\end{abstract}


\maketitle

\newcommand{\pos}{{\mathbf{r}}}
\newcommand{\dt}{{\Delta t}}
\newcommand{\dr}{{\Delta \pos}}
\newcommand{\vi}{{\mathbf{v}_i}}
\newcommand{\vm}{{v}}
\renewcommand{\u}{{\mathbf{u}}}
\newcommand{\eq}[1]{equation (\ref{#1})}
\newcommand{\Eq}[1]{Equation (\ref{#1})}
\newcommand{\fig}[1]{fig.~\ref{#1}}
\newcommand{\Fig}[1]{Fig.~\ref{#1}}
\newcommand{\teq}{\theta_{eq}}


From microfluidic technology to detergent design and ink-jet printing
it is important to investigate the way in which drops move across
surfaces. The dynamics of the drops will be affected by any chemical
heterogeneities on the
surface~\cite{degennes:85,lipowsky:01,darhuber:05}. Until recently
such disorder was usually regarded as undesirable. However with the
advent of microfabrication techniques it has become possible to
control the chemical patterning of a substrate down to nanoscale,
leading to the possibility of exploring how such patterning can
control, rather than disturb, drop motion.

A practical example where such an approach might be of use is in
ink-jet printing. Although ink-jet printers are widely available for
domestic use the quality of the images is still not sufficiently
robust to allow widespread industrial applications. The possibility of
replacing the traditional contact techniques with electronically
controlled template design, particularly for small print runs, is
highly desirable for both efficiency and cost.

\begin{figure}
\begin{center}
\begin{tabular}{cc}
(a) & (b) \\ \epsfig{file=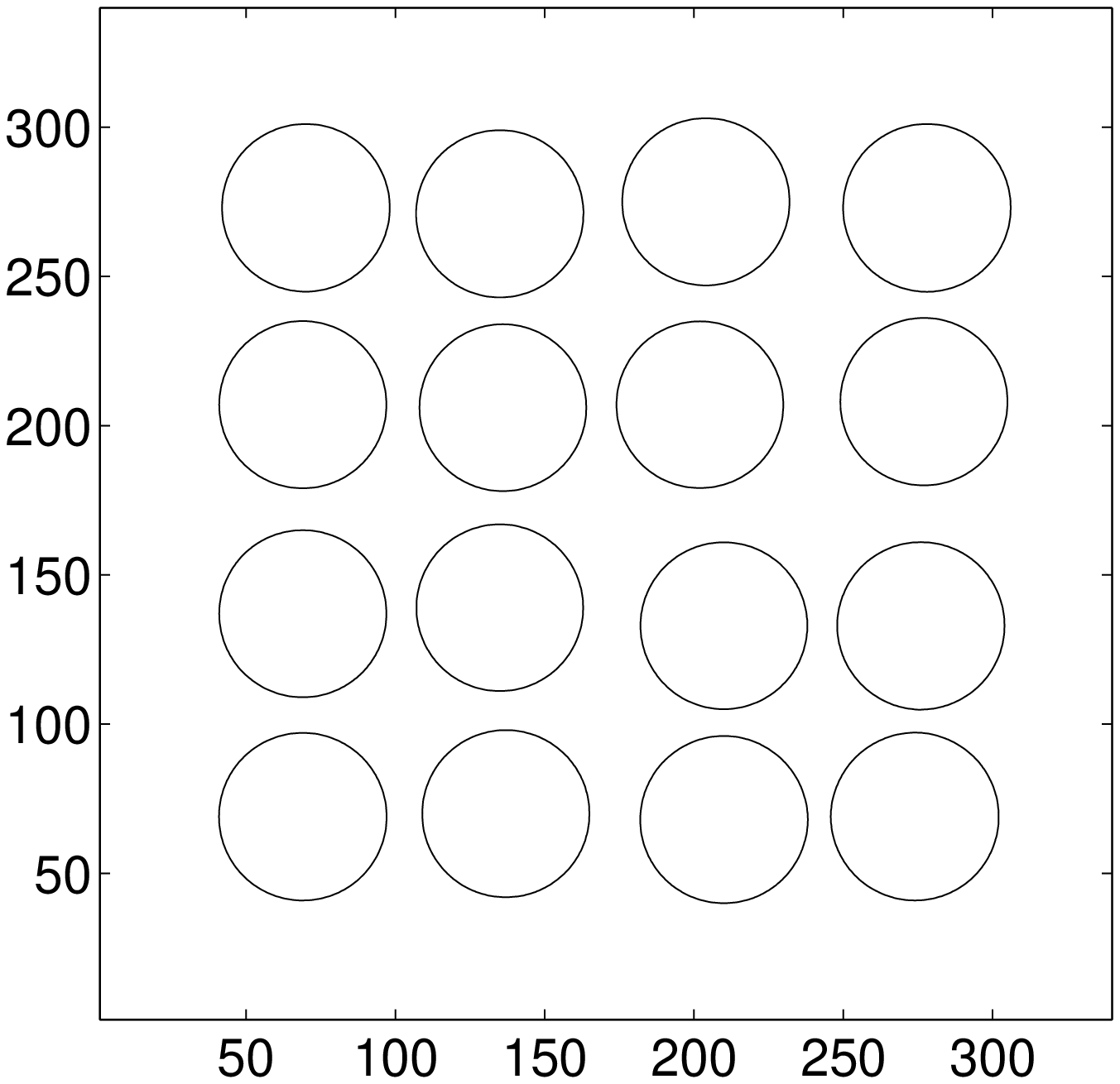,width=3.5cm} &
\epsfig{file=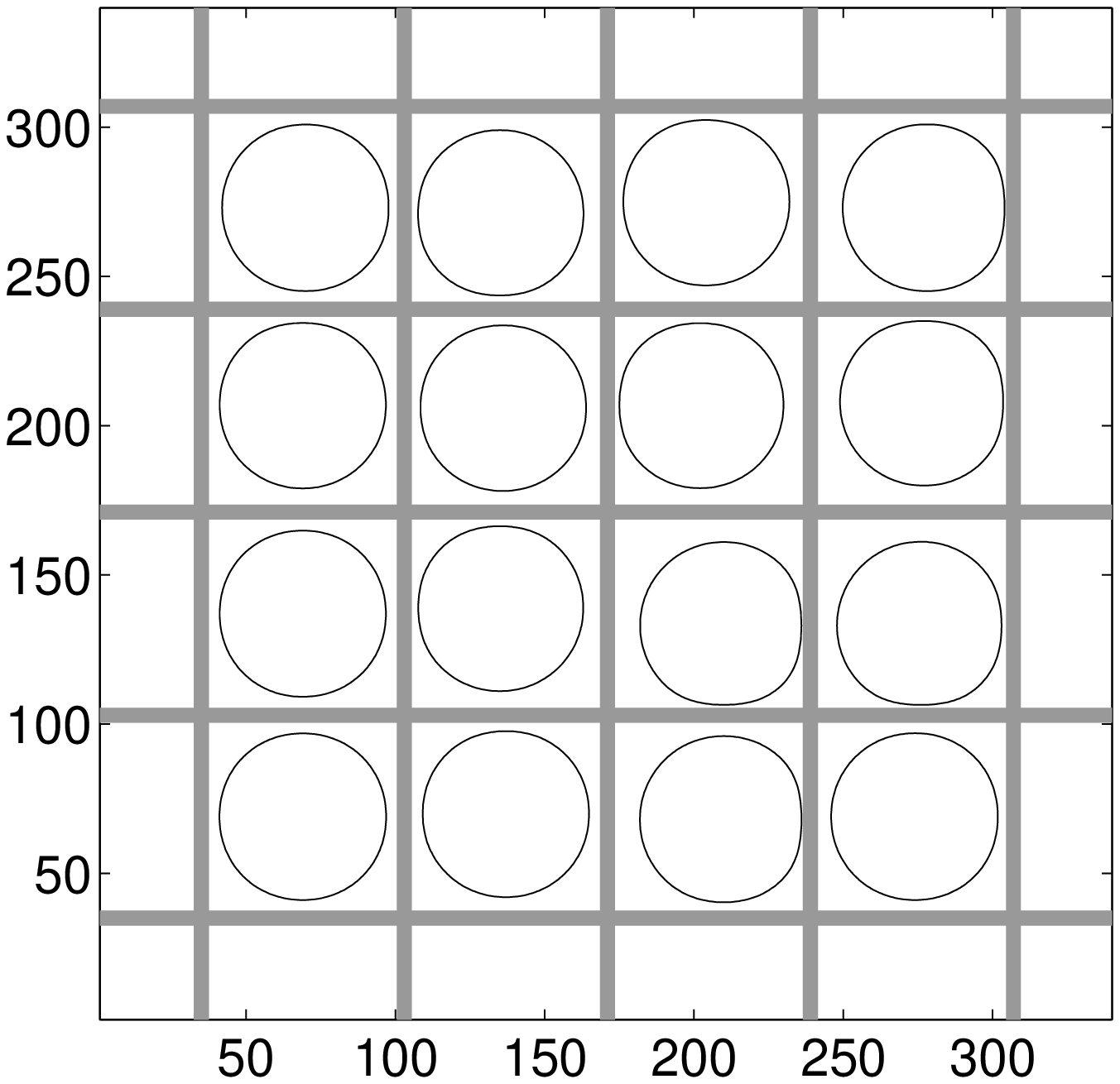,width=3.5cm} \\
\epsfig{file=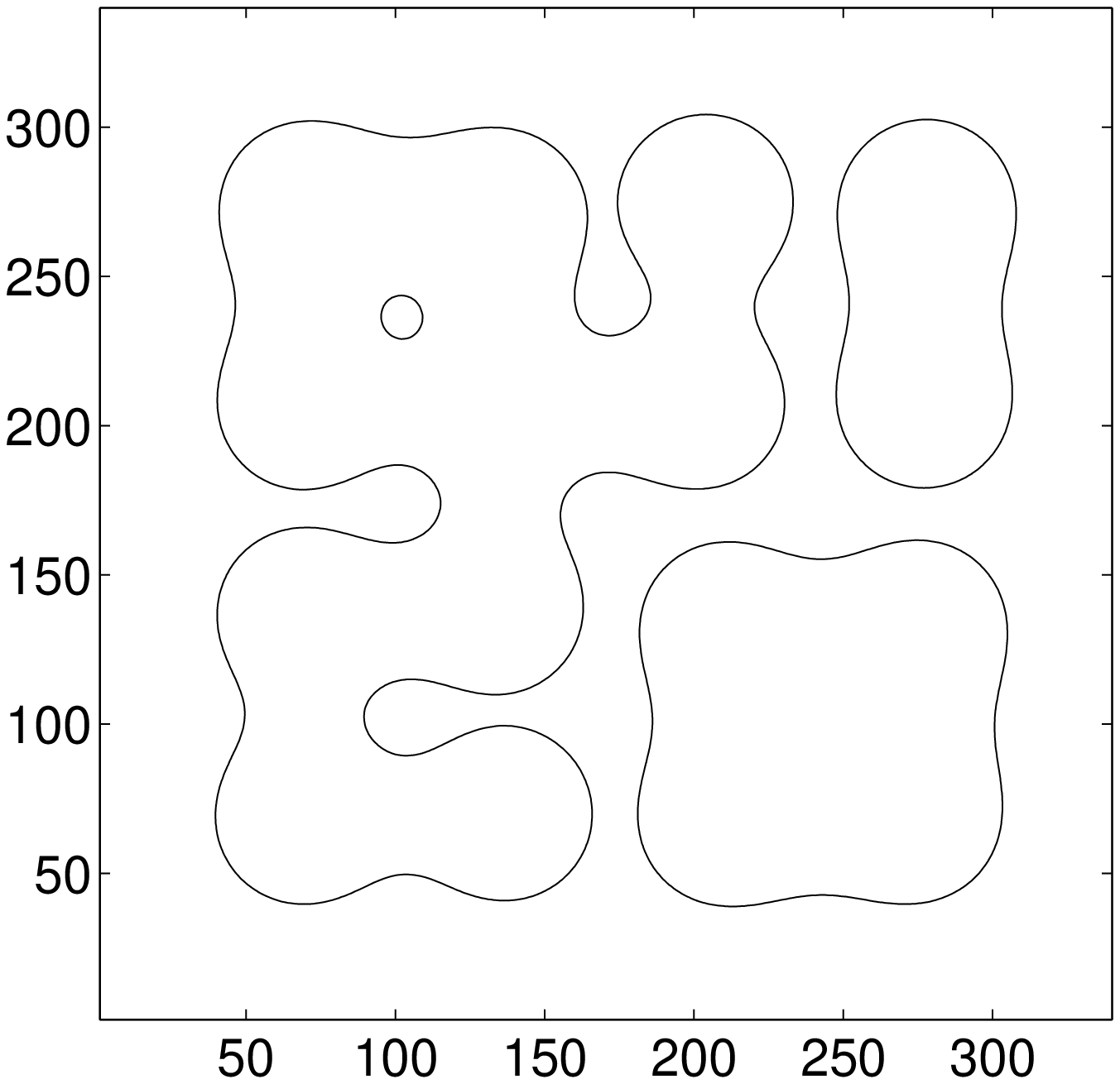,width=3.5cm} &
\epsfig{file=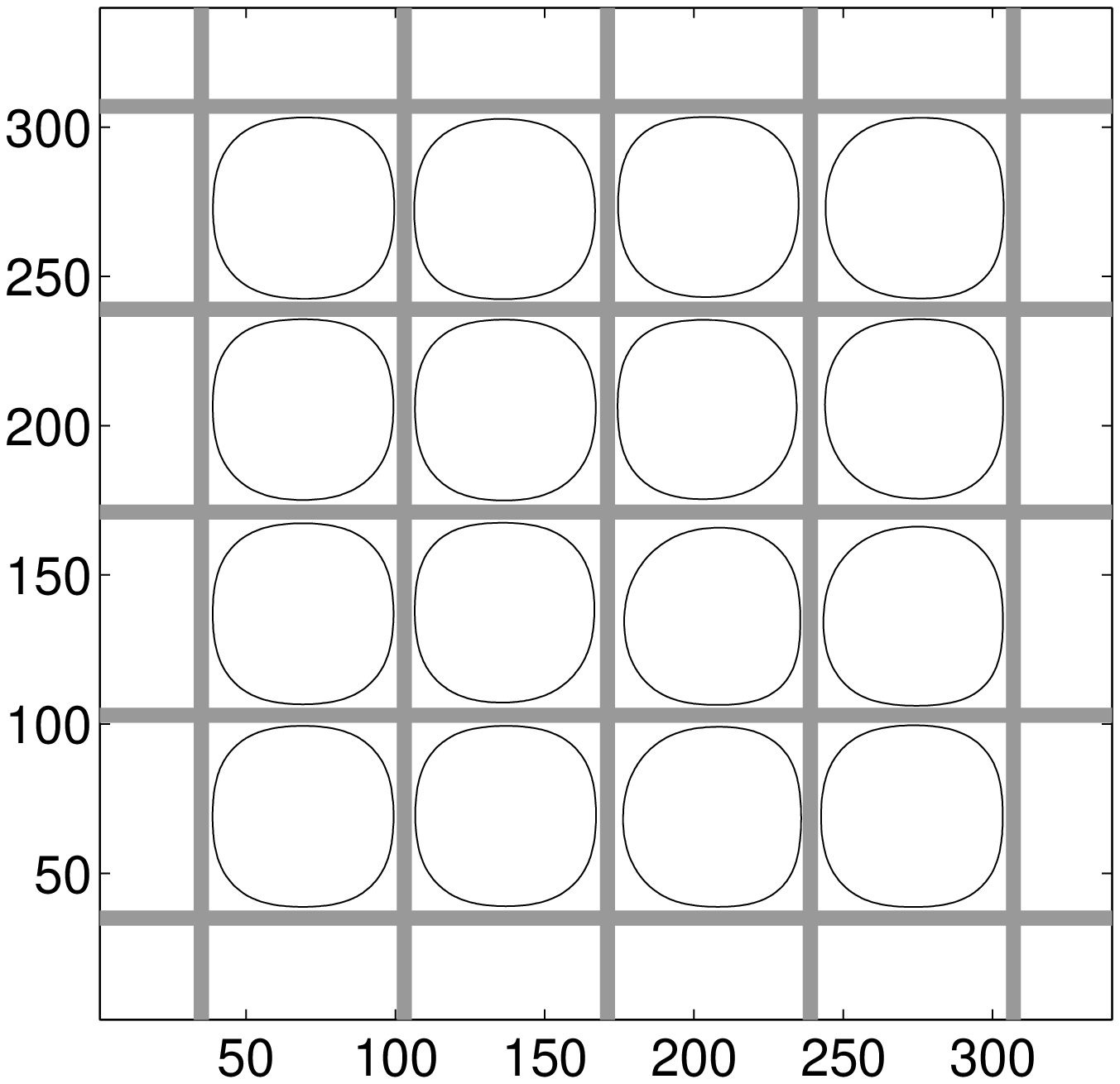,width=3.5cm} \\
\epsfig{file=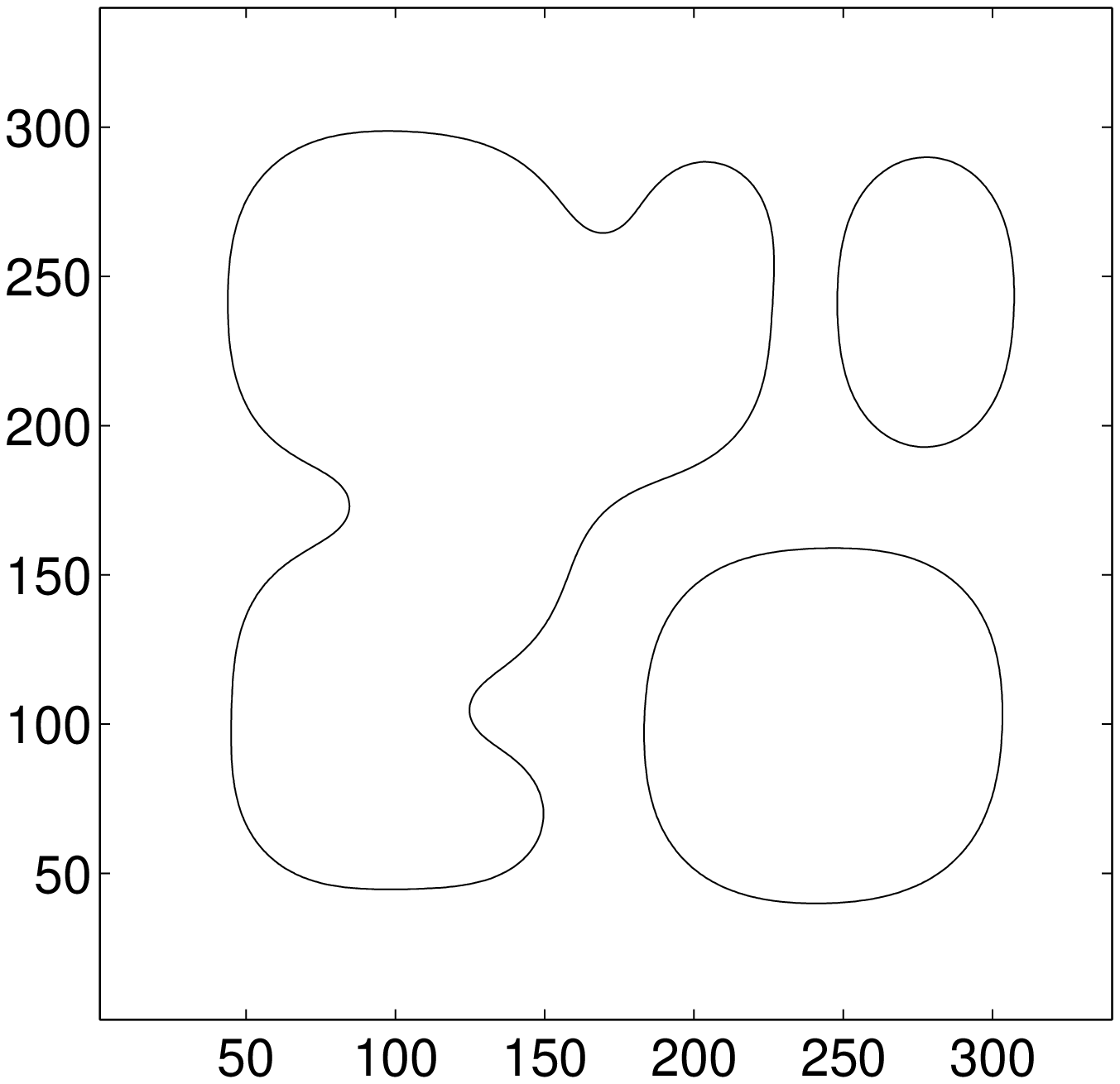,width=3.5cm} &
\epsfig{file=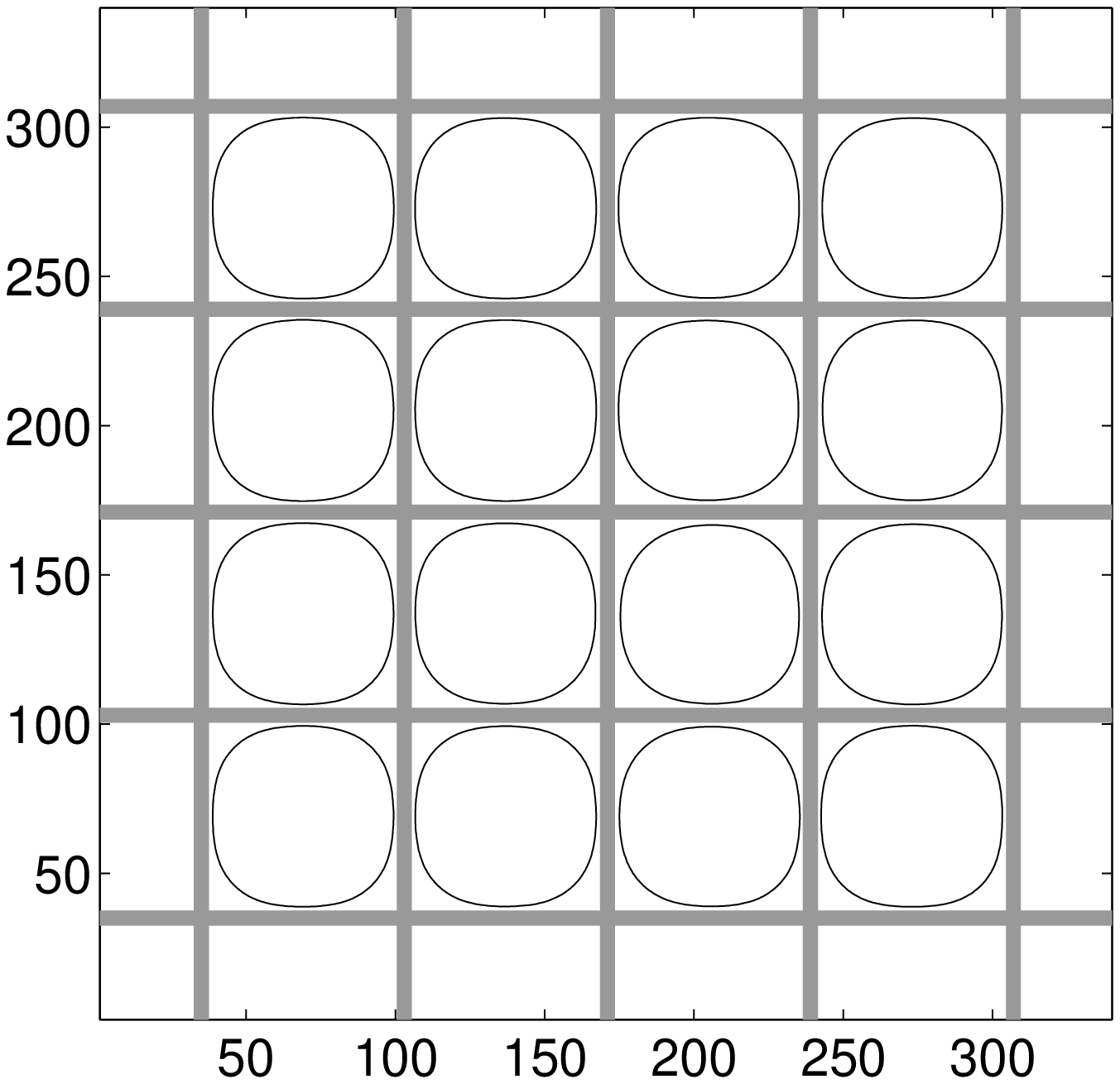,width=3.5cm} \\
\epsfig{file=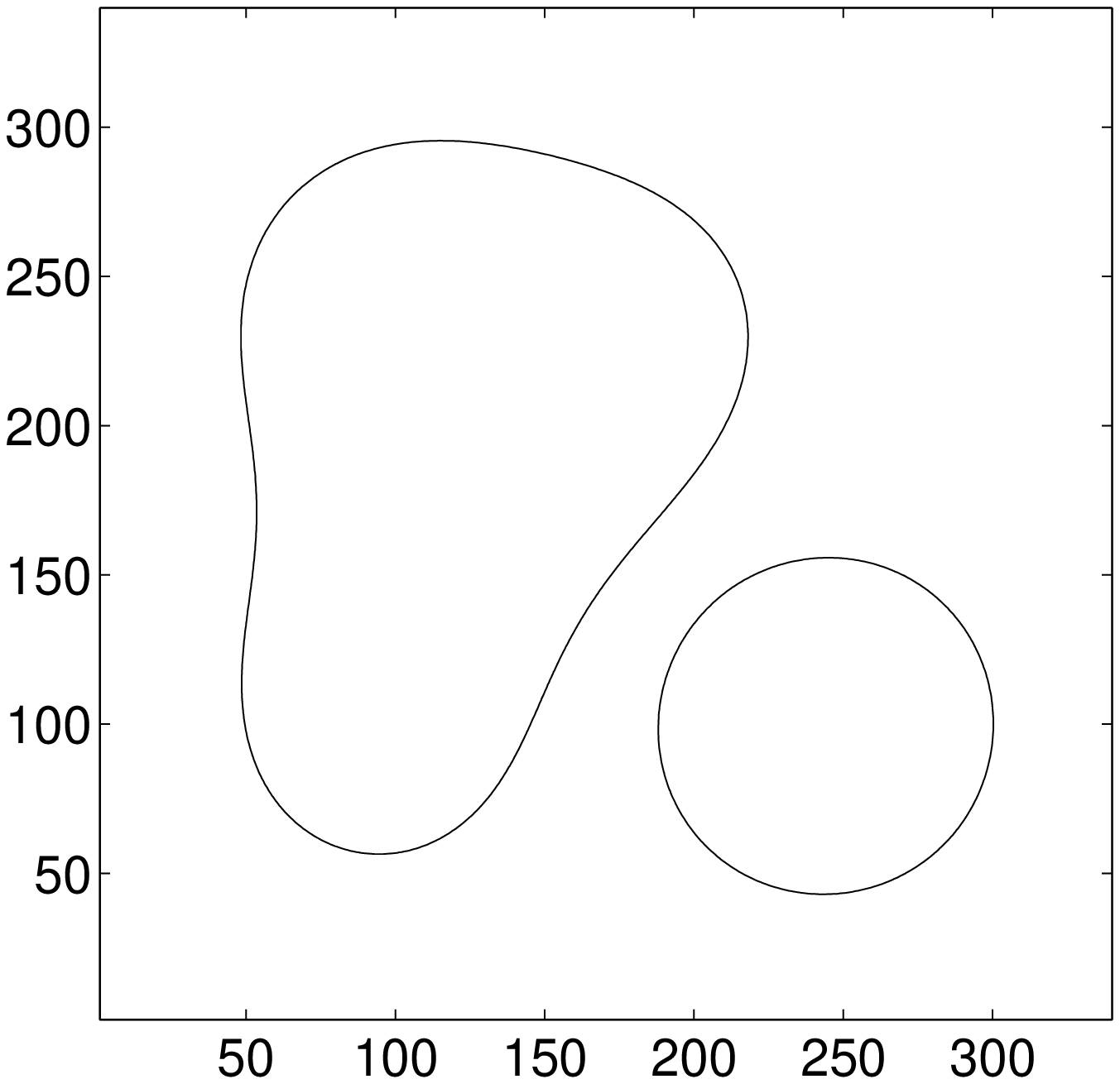,width=3.5cm} &
\epsfig{file=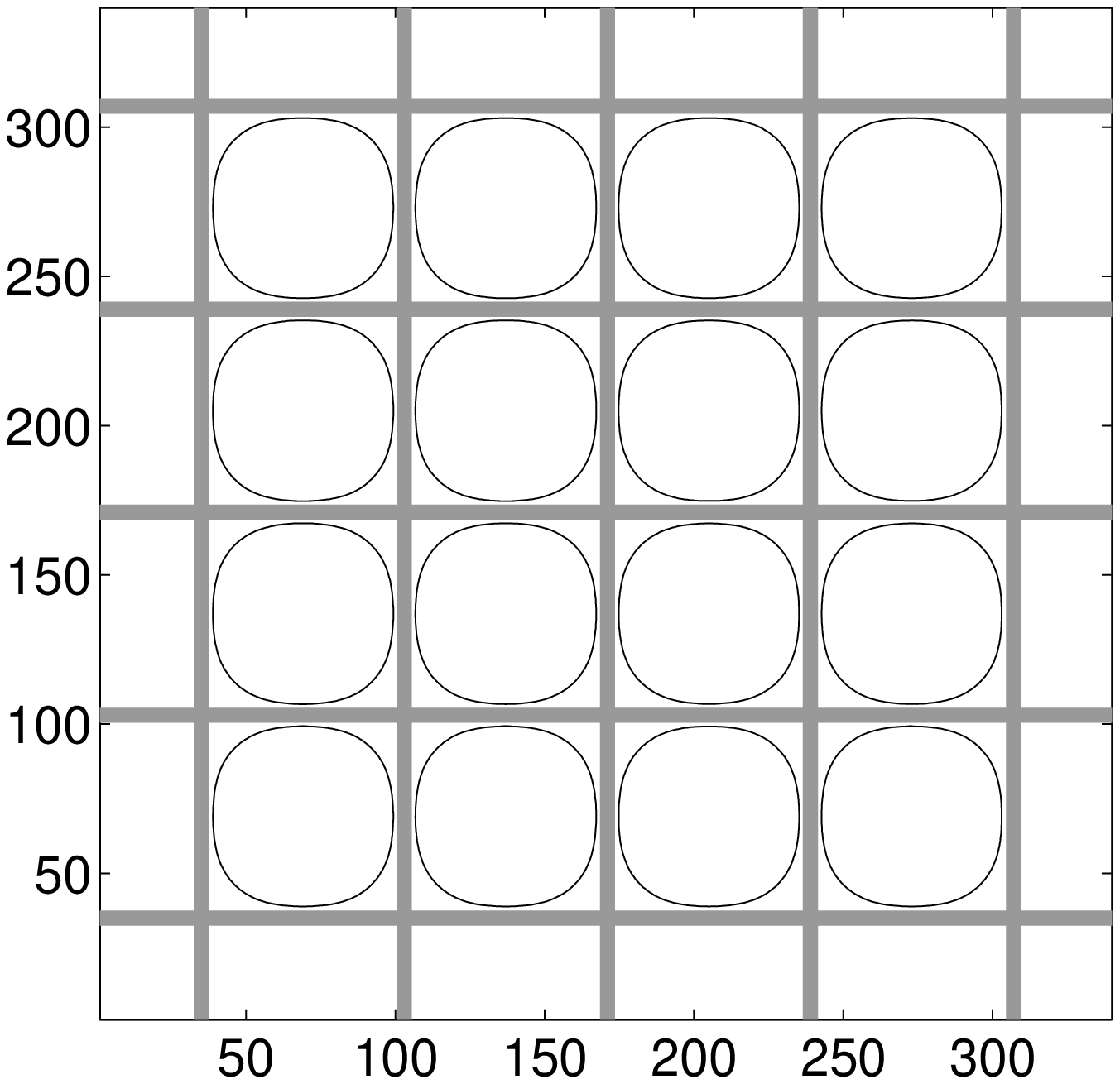,width=3.5cm} \\
\end{tabular}
\caption{Spreading of an array of drops which have small randomness 
in their initial impact point. (a) Substrate with contact angle
$\theta_{phi}=5^o$ on every sites. (b) $\theta_{phi}=5^o$ surface
patterned by $\theta_{pho}=65^o$ stripes. Stripes are regularly spaced
every $68 \mu$m and are $5 \mu$m wide. Figures are labeled in $\mu$m.}
\label{fig:mottle}
\end{center}
\end{figure}

\begin{figure}
\begin{center}
\vspace*{0.4cm}
\epsfig{file=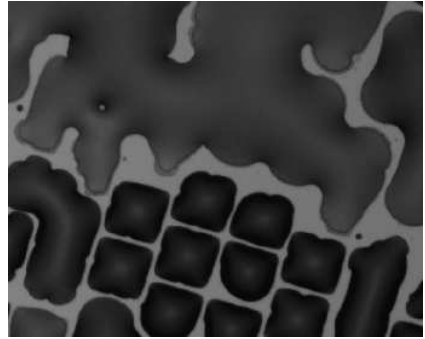,width=5.5cm}
\caption{Drops of radius $R=30 \mu$m which have been jetted onto a 
substrate and cured: (top) homogeneous and (bottom)
patterned. Hydrophilic and hydrophobic contact angles are
$\theta_{phi}=5^o$ and $\theta_{pho}=65^o$ respectively.}
\label{fig:mottleExpe}
\end{center}
\end{figure}

In the printed image a patch of colour is produced by jetting drops in
a regular, square array. The closer the drops the more intense the
colour of the patch appears to the eye. To achieve a solid colour the
aim is that drops jetted at a distance apart comparable to their
diameter should coalesce and form a uniform covering of ink. However,
in practice, randomness in the positions at which the drops land,
combined with surface imperfections, often lead to local coalescence
and the formation of large, irregular drops with areas of bare
substrate between them as shown in \fig{fig:mottle}(a) and the upper
part of \fig{fig:mottleExpe}. Such configurations are likely to lead
to poor image quality, called mottle~\cite{sandreuter:94}.

In an attempt to overcome this problem we demonstrate how using a
two-dimensional array of hydrophobic chemical stripes can be used to
control the equilibrium shape, the position and the dynamic pathway of
spreading drops. The hydrophobic stripes form barriers controlling the
drops and allowing their relative positions to be tuned. The behaviour
of single drops, and then an array of drops on the patterned surfaces,
is explored both by solving the hydrodynamics equations of motion by
means of a lattice Boltzmann algorithm and by performing suitable
experiments.

In the numerical modelling we consider a liquid-gas system of density
$n(\pos)$ and volume $V$. The surface of the substrate is denoted by
$S$. The equilibrium properties of the drop are described by the free
energy
\begin{equation}
\Psi = \int_V \left( \psi_b(n) + \frac{\kappa}{2} \left( \partial_\alpha n
  \right)^2 \right) dV + \int_S \psi_c(n) \; dS
\label{eq:freeE}
\end{equation}
where Einstein notation is understood for the Cartesian label $\alpha$
and where $\psi_b(n)$ is the free energy in the bulk which we take to
have a Van der Waals form. The derivative term in \eq{eq:freeE} models
the free energy associated with density gradients at an
interface. $\kappa$ is related to the surface tension. Following
Cahn~\cite{cahn:77} we choose $\psi_c(n_s)=-\phi_1 n_s$, where $n_s$
denotes the density at the surface, to control the wetting properties
at the fluid.

We use a lattice Boltzmann model to solve the Navier-Stokes equations
of this liquid-gas system 
\begin{eqnarray}
\partial_t (n u_\alpha) + \partial_\beta (nu_\alpha u_\beta) & = & 
  - \partial_\beta P_{\alpha\beta} + \nonumber \\
  \nu \partial_\beta [ n (\partial_\beta u_\alpha & + & 
                              \partial_\alpha u_\beta + 
                              \delta_{\alpha\beta} \partial_\gamma u_\gamma) 
                       ], \nonumber \\
\partial_t n + \partial_\alpha(n u_\alpha) & = & 0
\label{eq:ns} 
\end{eqnarray}
%
%
%
where $\mathbf{u}(\pos)$ is the fluid velocity and $\nu$ the kinematic
viscosity. The pressure tensor $P_{\alpha\beta}$ incorporates
information about the free energy~\cite{anderson:98}. Details of the
numerical algorithm can be found in~\cite{dupuis:04e}. No slip
boundary conditions on the velocity are set on all surfaces. In what
follows we consider ink droplets of viscosity $\eta=2.5\cdot 10^{-2}$
kgm$^{-1}$s$^{-1}$ and surface tension $\sigma=2 \cdot 10^{-2}$
Nm$^{-1}$.

\begin{figure}[htbp]
\begin{center}
\begin{tabular}{cc}
(a) & (b) \\
\epsfig{file=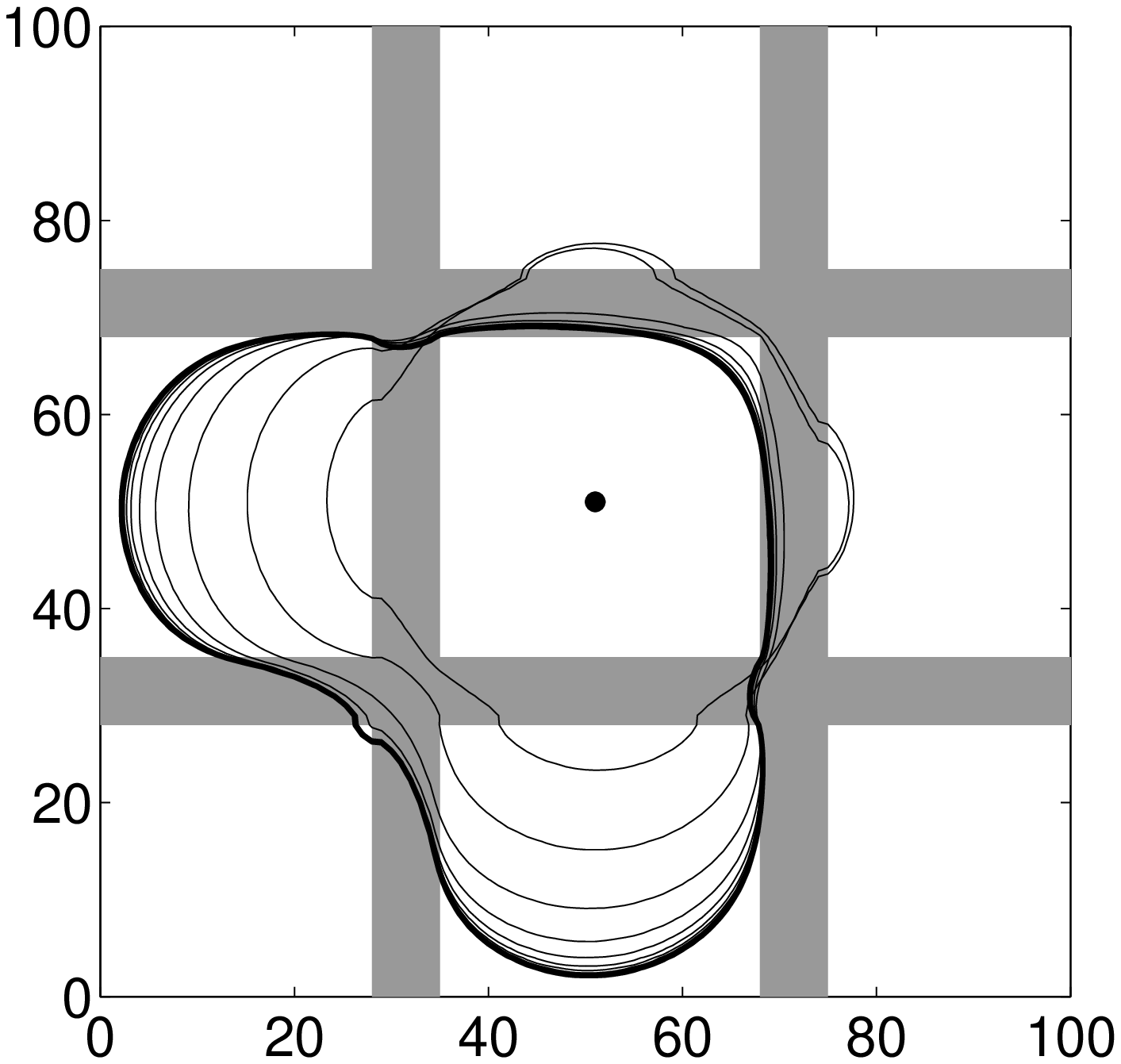,width=4cm} &
  \epsfig{file=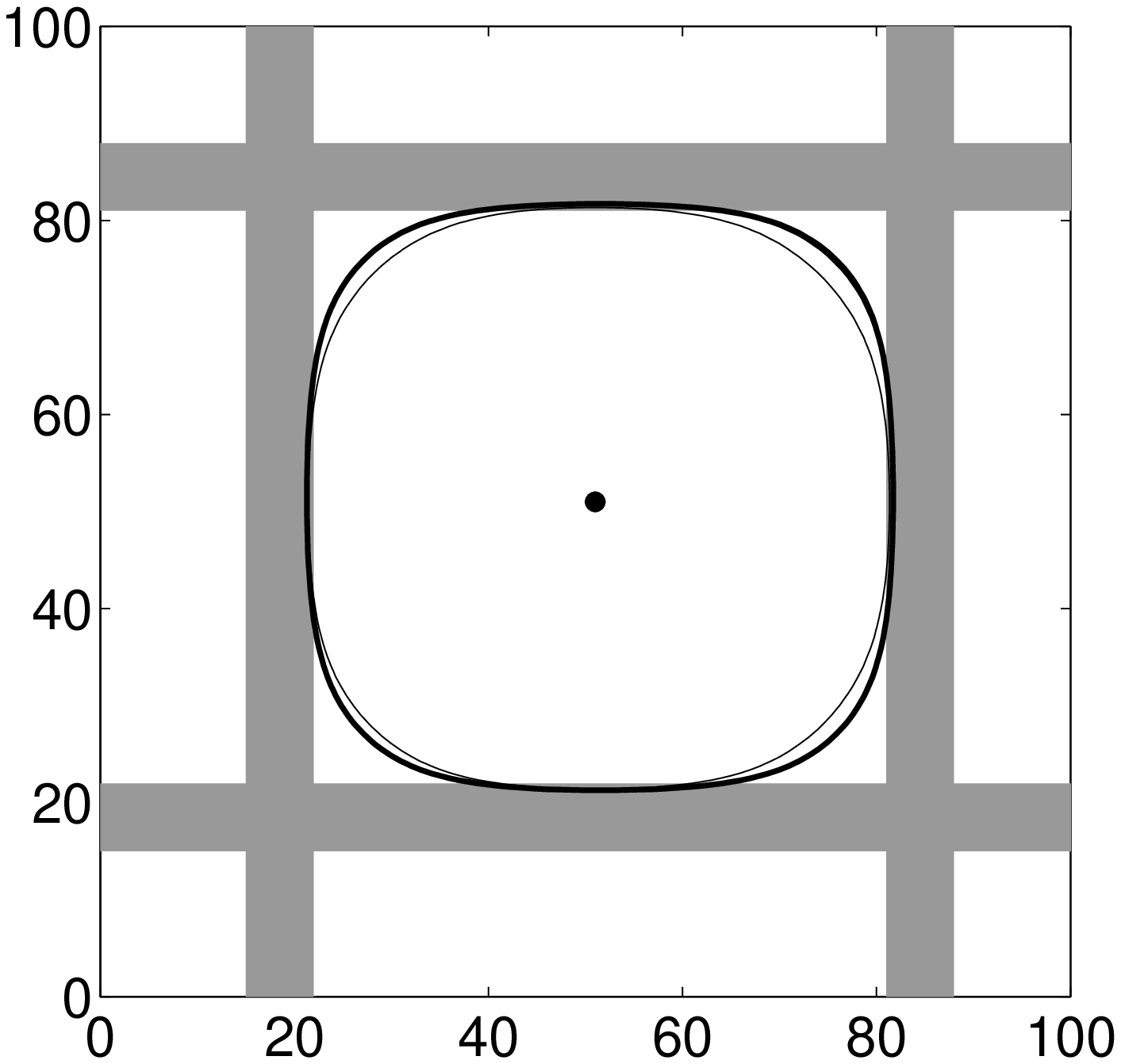,width=4cm} \\
(c) & (d) \\
  \epsfig{file=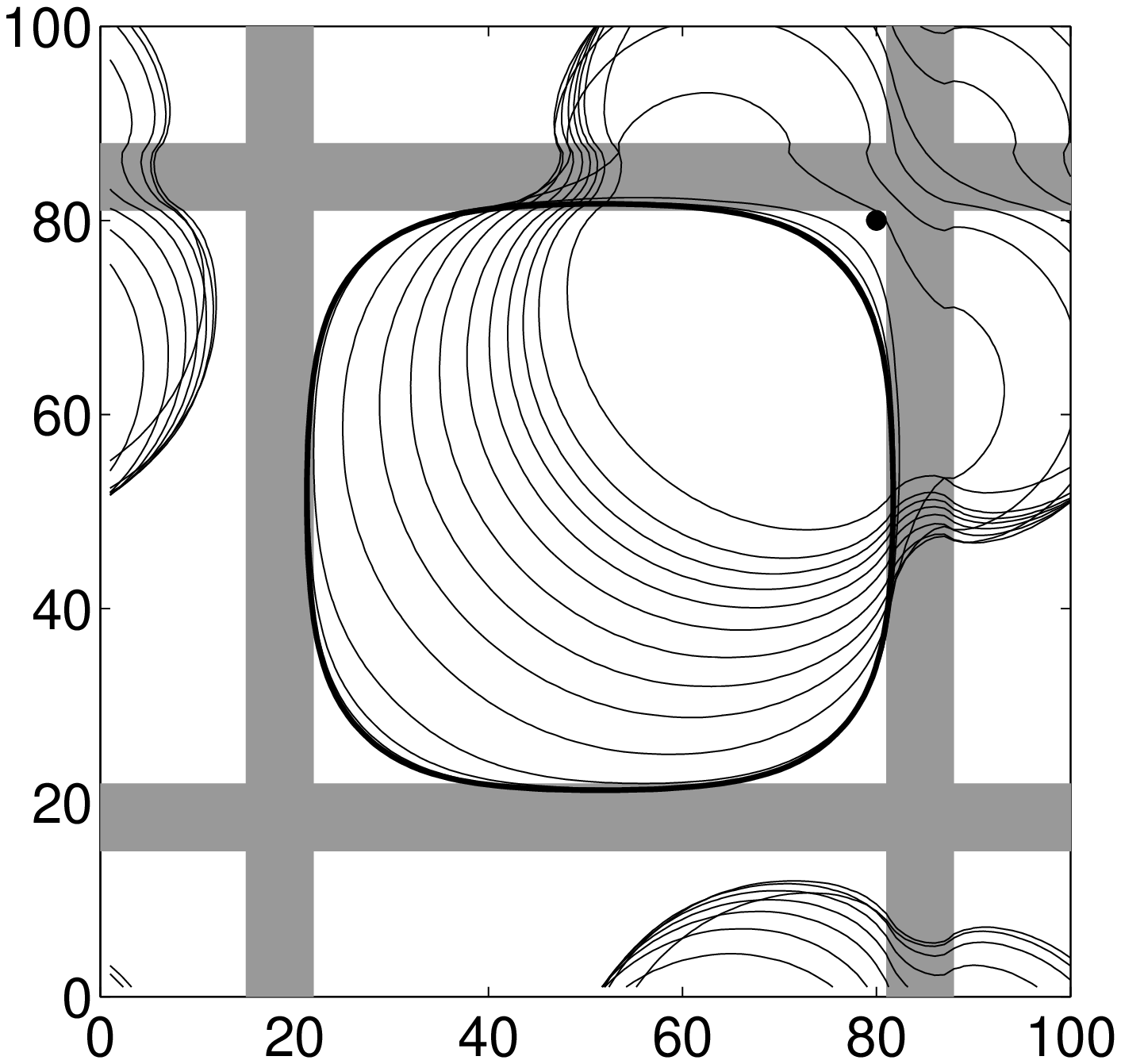,width=4cm} &
  \epsfig{file=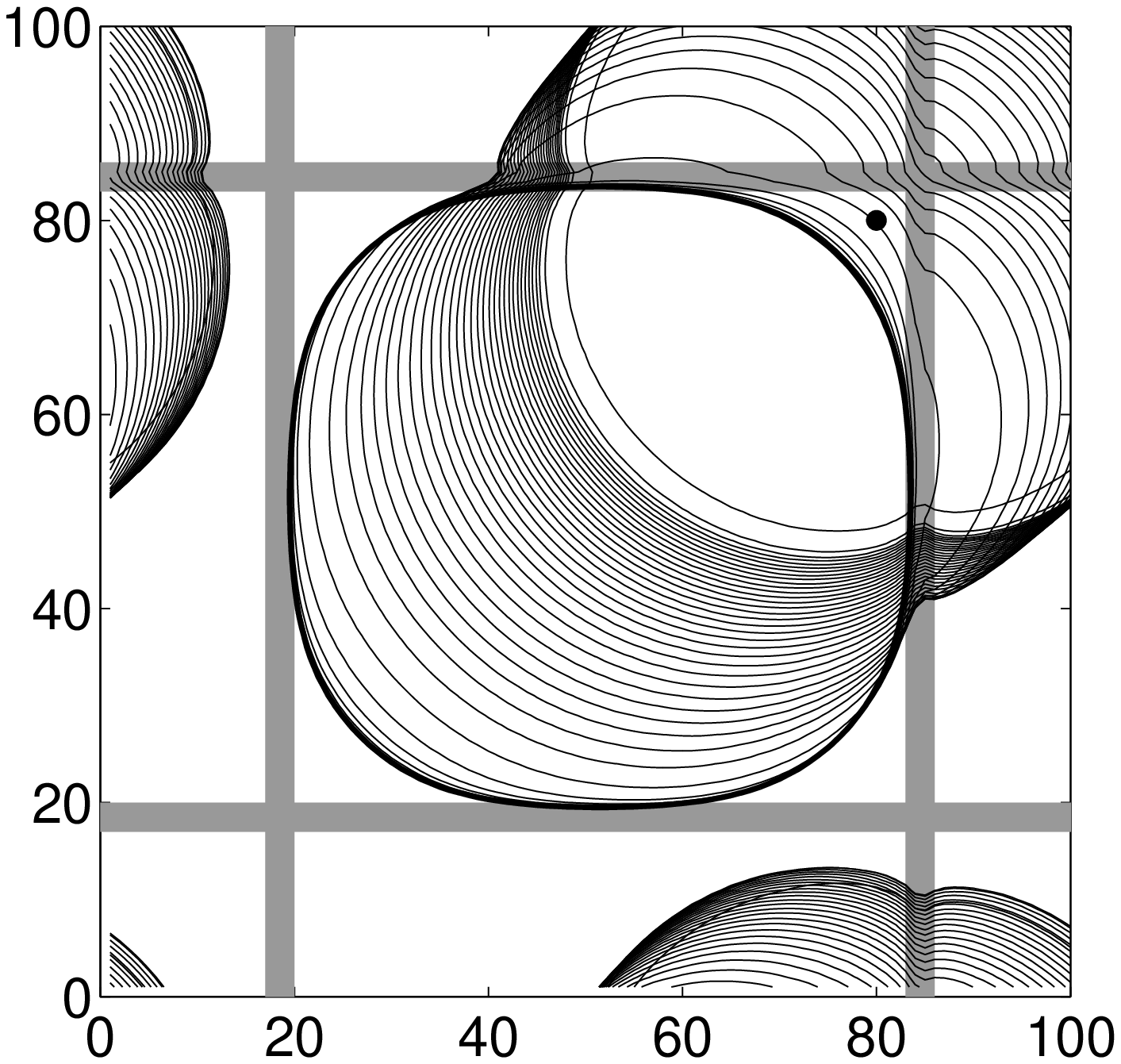,width=4cm} \\
\end{tabular}
\caption{Time evolution of drops jetted onto substrates patterned 
by grids.  Hydrophobic and hydrophilic areas are dark grey stripes
($\theta_{pho} = 65^o$) and white areas ($\theta_{phi} = 5^o$)
respectively. Black dots are impact points. Thin lines show how the
drop spreads in time and thick lines show the final shape. The droplet
radius is $15 \mu m$. Equilibrium is reached after (a-c) $2$ms, (d)
$2.7$ms. (a) Hydrophobic stripes width $w=6 \mu$m and distance between
centre of stripes $d=40 \mu$m. The square is too small and the drop
escapes. (b) $w=6 \mu$m and $d=66 \mu$m. The drop is confined.  (c)
$w=6 \mu$m and $d=66 \mu$m. The drop is pulled back and confined
despite landing at the corner of a square. (d) $w=4 \mu$m and $d=66
\mu$m. Thinner stripes slow down the confinement process. Figures 
are labeled in $\mu$m.}
\label{fig:squares}
\end{center}
\end{figure}

The final state of a drop of liquid placed on a solid surface depends
on the wetting properties of the surface~\cite{degennes:85}. These are
best characterised by the contact angle $\theta$ (which can be
controlled in our simulations by choosing an appropriate
$\phi_1$). Drops prefer to lie on hydrophilic portions of the
substrate, i.e. areas with smaller $\theta$. Substrates can be
chemically patterned with for example stripes, spots or grids of
different material to create areas where the contact angle
differs. Here we consider grids of relatively hydrophobic stripes with
contact angle $\theta_{pho}=65^o$ on a substrate which otherwise has a
contact angle $\theta_{phi}=5^o$. \Fig{fig:squares} shows the
behaviour of drops of radius $R=15 \mu$m which are placed on such
surfaces so that they are initially just touching the surface at the
point marked with a black dot. Thin lines on the figures show how the
drop spreads in time and thick lines show the final shape. The figures
compare different spacing between, and width of, the grid lines and
different impact points of the drop.

We first consider, in \fig{fig:squares}(a), stripes of width $6 \mu$m
spaced by $40 \mu$m. In this case the hydrophilic area is too small
compared to the drop volume to confine the drop within a single square
(the south-west drift is due to the impact point being set slightly
off centre). \Fig{fig:squares}(b) shows that increasing the spacing
between the stripes to $66 \mu$m creates an hydrophilic region big
enough to confine the drop. Surprisingly this is true for any point of
impact within the hydrophilic square as illustrated in
\fig{fig:squares}(c) where the drops lands in the corner of the
square.

The confinement occurs because the surface tension penalty, which
results from the final shape of the drop being non-spherical, is
outweighted by the advantage of not having to lie on the hydrophobic
regions of the surface. For thinner hydrophobic stripes, $4 \mu$m,
shown in \fig{fig:squares}(d), the free energy penalty is smaller and
hence the driving force for confinement is less and the drops takes
an extra $0.7$ms to be pulled back into its original square.

\begin{figure}
\begin{center}
\epsfig{file=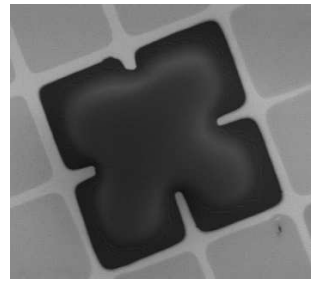,width=4cm}
\end{center}
\caption{An ink drop spread over a chemically patterned surface. Light
and dark grey areas correspond to hydrophobic ($\theta_{pho}=65^o$)
and hydrophilic ($\theta_{phi}=5^o$) regions respectively. The drop
radius is $30 \mu$m.}
\label{fig:expeSquares}
\end{figure}

\Fig{fig:expeSquares} shows an experimental ink drop of radius $R=30
\mu$m jetted onto a chemically patterned surface. Technical
details for designing such surfaces can be found
in~\cite{leopoldes:03}. The surface comprises $60 \mu$m$\times 60
\mu$m hydrophilic ($\theta_{phi}=5^o$) squares separated by $5 \mu$m 
wide hydrophobic ($\theta_{pho}=65^o$) stripes.

The drop shows strong confinement (but within four squares, because
the relative sizes of drop and grid are larger than in the
simulation). The equilibrium shape reflects the underlying
patterning. \Fig{fig:expeSquares} also shows non-wetted regions along
the inner hydrophobic stripes.
 


We now turn to consider the behaviour of an array of drops, in
approximate registry with the hydrophobic grid. Our aim is to show how
chemical patterning can be used to address the problem of mottle,
uneven droplet coalescence, that can severely limit image quality in
ink jet printing.

We consider two surfaces (a) and (b) on which an array of $15 \mu$m
radius drops are jetted at $3 $ms$^{-1}$. Surface (a) is homogeneous
with an equilibrium contact angle $\theta_{phi}=5^o$ whereas surface
(b) is patterned by vertical and horizontal stripes of contact angle
$\theta_{pho}=65^o$. Stripes are regularly spaced every $68 \mu$m and
are $5 \mu$m wide. In the areas between the stripes, the contact angle
$\theta_{phi}=5^o$.

The drops are jetted so as to hit the surface at approximately the
middle of each hydrophilic square. Randomness in the position of
impact is however set by adding a $\pm 5\mu$m
noise. \Fig{fig:mottle} shows the spreading of drops with and without
chemical patterning.

\Fig{fig:mottle}(a) shows that, on substrate (a), the randomness of 
the drop impact points produces an uneven and complicated pattern. The
drops which land slightly closer together coalesce first and
immediately start to dewet the surrounding substrate as they minimise
their free energy by forming a larger spherical drop. This process
results in larger, randomly spaced, isolated drops, with undesirable
areas of bare substrate between them.


In contrast, on substrate (b), the evolution starting from {\it
identical} initial condition shows that a hydrophobic grid can control
the final drop position. Drops do not coalesce and form a regular
array. The coverage is higher and is likely to lead to a better image
quality.

An experiment presenting a similar situation is shown in
\fig{fig:mottleExpe}. The ink drops have a radius $R=30 \mu$m and they
are jetted in a $50 \mu$m$\times 50 \mu$m array. In the upper part of
the figure there is no hydrophobic grid and a mottled final drop
configuration is observed. The configuration is equivalent to the
second frame in \fig{fig:mottle}(a) because the drops are cured before
reaching their final equilibrium state. There are also likely to be
surface heterogeneities which may pin the drop.

This is no longer the case when the underlying surface is
patterned. The lower part of \fig{fig:mottleExpe} carries hydrophobic
stripes of width $5 \mu$m forming squares of side $40 \mu$m. The drops
now form a more regular array determined by the grid. We note that each
drop covers four grid squares, as the drop radius to square side length
ratio is larger than in the simulations.

%
%
In this letter we have demonstrated, both numerically and
experimentally, that the chemical patterning of a substrate is
surprisingly effective in controlling drop positions on a
substrate. In particular the tendency of an array of drops with small
randomness in their points of impact at the substrate to mottle can be
controlled. It may be possible to exploit this technique to improve
the quality of ink-jet images.
%
\section*{Acknowledgment}
This work forms part of the IMAGE-IN project which is funded by the
European Community through a Framework 5 grant (contract number:
GRD1-CT-2002-00663).


\begin{thebibliography}{1}

\bibitem{degennes:85}
P.G. de~Gennes.
\newblock Wetting: statics and dynamics.
\newblock {\em Rev. {M}od. {P}hys.}, 57:827, 1985.

\bibitem{lipowsky:01}
R.~Lipowsky.
\newblock Morphological wetting transitions at chemically structured surfaces.
\newblock {\em Curr. {O}pin. {C}olloid {I}nterface {S}ci.}, 6:40, 2001.

\bibitem{darhuber:05}
A.A. Darhuber and S.M. Troian.
\newblock Principles of microfluidic actuation by modulation of surface
  stresses.
\newblock {\em Ann. {R}ev. {F}luid {M}ech.}, 37:425, 2005.

\bibitem{sandreuter:94}
N.P. Sandreuter.
\newblock Predicting print mottle - {A} method of differentiating between 3
  types of mottle.
\newblock {\em Tappi {J}ournal}, 77:173, 1994.

\bibitem{cahn:77}
J.W. Cahn.
\newblock Critical point wetting.
\newblock {\em J. {C}hem. {P}hys.}, 66:3667, 1977.

\bibitem{anderson:98}
D.M. Anderson, G.B. McFadden, and A.A. Wheeler.
\newblock Diffuse-interface methods in fluid mechanics.
\newblock {\em Annu. {R}ev. {F}luid {M}ech.}, 30:139, 1998.

\bibitem{dupuis:04e}
A.~Dupuis and J.M. Yeomans.
\newblock Modelling droplets on superhydrophobic surfaces: equilibrium states
  and transitions.
\newblock {\em Langmuir}, 21:2624, 2005.

\bibitem{leopoldes:03}
J.~L{\'e}opold{\`e}s, A.~Dupuis, D.G. Bucknall, and J.M. Yeomans.
\newblock Jetting micron-scale droplets onto chemically heterogeneous surfaces.
\newblock {\em Langmuir}, 19:9818, 2003.

\end{thebibliography}
\end{document}